\documentclass[english]{lni}
\begin{document}
%%% Mehrere Autoren werden durch \and voneinander getrennt.
%%% Die Fußnote enthält die Adresse sowie eine E-Mail-Adresse.
%%% Das optionale Argument (sofern angegeben) wird für die Kopfzeile verwendet.
\title[]{LLM-Based Agentic Systems for Software Engineering: Challenges and Opportunities}
 \author[1,2]{Yongjian Tang}{yongjian.tang@siemens.com}{0009-0008-8670-6843}
 \author[1,2]{Thomas Runkler}{thomas.runkler@siemens.com}{0000-0002-5465-198X}
 \affil[1]{Siemens AG\\Garching near Munich\\Germany}
 \affil[2]{Technical University of Munich\\Garching near Munich\\Germany}
 %\affil[1]{Siemens AG\\Friedrich-Ludwig-Bauer-Straße 3\\ 85748 Garching near Munich\\Germany}
 %\affil[2]{Technical University of Munich\\Boltzmannstraße 15\\85748  Garching near Munich\\Germany}
\maketitle

\begin{abstract}
% Despite recent advancements, Large Language Models (LLMs) often fall short in tackling complex Software Engineering (SE) tasks. 
Despite recent advancements in Large Language Models (LLMs), complex Software Engineering (SE) tasks require more collaborative and specialized approaches. This concept paper systematically reviews the emerging paradigm of LLM-based multi-agent systems, examining their applications across the Software Development Life Cycle (SDLC), from requirements engineering and code generation to static code checking, testing, and debugging. We delve into a wide range of topics such as language model selection, SE evaluation benchmarks, state-of-the-art agentic frameworks and communication protocols. Furthermore, we identify key challenges and outline future research opportunities, with a focus on multi-agent orchestration, human-agent coordination, computational cost optimization, and effective data collection. This work aims to provide researchers and practitioners with valuable insights into the current forefront landscape of agentic systems within the software engineering domain.
\end{abstract}

%%% Beginn des Artikeltexts

\section{Introduction}
%\subsection{Motivation}
Large Language Models (LLMs)~\cite{zhao2024surveylargelanguagemodels} have shown promising potential to transform the Software Development Lifecycle (SDLC) .
Their exceptional capabilities in text understanding, knowledge inference, and reasoning have opened new avenues for addressing complex software engineering tasks. %have provided us with more problem-solving possibilities. 
Both academia and industry have explored a wide range of techniques to address various Software Engineering (SE) use cases, such as prompt engineering \cite{sahoo2024systematic, tang2024fsponer}, in-context learning \cite{dong2024survey, tang2025few, nishtha}, parameter-efficient fine-tuning \cite{hanparameter}, and retrieval augmented generation \cite{gao2023retrieval}. Each approach offers distinct advantages for tailoring LLM capabilities to specific SE tasks. %, from code generation and bug detection to documentation and requirements analysis.

Building on these foundational techniques, LLM-based Multi-agent systems have emerged as a promising paradigm to address the growing complexity of modern software development. These systems can orchestrate multiple specialized agents, integrate diverse development tools, access external knowledge bases, and leverage domain-specific resources to provide comprehensive, end-to-end solutions for real-world SE challenges.

%In this work, we provide a systematic overview of LLM-based multi-agent systems in software engineering, structured as follows: 
This work is structured as follows: 
Section 2 introduces the backgrounds and foundations of language models and agentic systems. Section 3 presents a systematic literature review and investigates existing multi-agent approaches across requirements engineering, code generation, static code checking, testing, and debugging. Section 4 outlines our methodological approaches and research agenda, covering cutting-edge language models, SE benchmarks, and agentic frameworks. Section 5 proposes future challenges and opportunities, with a focus on agent capability enhancement, human-agent coordination, data collection, and cost optimization.

\section{Backgrounds and Foundations}
\subsection{Transformers and Language Models}
With the advent of transformer architecture~\cite{vaswani2017attention}, language modeling has received significant attention. 
The highly parallelizable architecture not only scales well with massive data and computing power but also effectively captures long-range dependencies and contextual relationships in text. 
BERT~\cite{Devlin2019BERTPO} initially stands out among all models. The semantic representations obtained in pre-training make it approachable for a broad spectrum of downstream tasks in specific domains. This ”pre-training and fine-tuning” mode has inspired numerous follow-up works.
Meanwhile, the research community continues to improve the performance of language models by scaling up their sizes. Compared with smaller counterparts, large-scale models demonstrate unseen emergent abilities~\cite{wei2022emergent} and have provided us with more problem-solving possibilities, such as prompt engineering, in-context learning, and step-by-step reasoning. 

\subsection{LLM-based Agentic Systems}
Facilitated by these LLMs, traditional agent-based systems evolved significantly. Such LLM-based agentic systems are now capable of handling fuzzy textual input, learning from demonstration examples, generating more contextually relevant responses, thus providing more coherent and meaningful interaction with users.

The main advantages of such collaborative agentic systems can be summarized as follows: 

\begin{itemize}
    \item   
    Each agent can be specialized for its specific task. For instance, one agent can focus on code generation, another on testing, another on documentation, etc. Specialization allows each agent to be optimized for its specific task, leading to superior performance and efficiency. 

    \item
    Multi-agent systems are inherently modular. If a particular agent needs a new capability, it can be upgraded independently through fine-tuning or prompting without affecting the entire system. This modularity facilitates better maintenance and easier integration of new technologies.
    
    \item 
    Multi-agent systems benefit from collaborative problem-solving capabilities, with each agent contributing a unique perspective, resulting in more versatile solutions.

    \item  
    Multi-agent systems excel in using tools and other resources, leading to more efficient and effective outcomes than those achieved by a single large language model.
    
    \item 
    Multi-agent systems can perform tasks in parallel, speeding up the software development.

\end{itemize}

\section{Literature Review -- Agents for SE }
%\subsection{LLM-based Agentic Systems for SE}
A wide range of studies have explored the integration of LLM-based multi-agent systems across different stages of SDLC. We present a concise review of selected works, categorizing them into five pivotal fields of software engineering -- requirements engineering, code generation, static code checking, testing, and debugging. 

\subsection{Requirements Engineering}
Requirements engineering focuses on defining and managing system requirements. This stage consists of several interconnected processes: elicitation, modeling, specification, analysis, and validation. These processes work collectively to ensure that requirements align with stakeholders' needs while adhering to pre-defined quality standards. 
% A promising approach to ensure traceability and management is requirement classification, which provides a structured way to organize requirements based on their characteristics and attributes. Several works~\cite{norbert,prcbert} fine-tune language models to classify software requirements based on their characteristics and attributes. A recent study~\cite{el2023ai} directly prompts ChatGPT to categorize requirements in few-shot settings, achieving performance comparable to fine-tuned models. 

Many LLM-based agents have been proposed to address the challenges in this stage. For instance, a multi-agent system~\cite{sami2024early} is built to generate and prioritize use stories from the initial project description, effectively identifying the core features of the application and thus enhancing the requirement elicitation and analysis. Furthermore, a framework called SpecGen~\cite{ma2024specgen} divides specification generation into two phases. The first phase employs a
conversational approach, guiding the LLM to generate relevant specifications for a given program. The second phase applies four mutation operators to the model-generated specifications and selects verifiable specifications from the mutated ones through a heuristic selection strategy. 
MARE~\cite{jin2024mare}, another multi-agent system covering several phases of requirements engineering, performs nine actions to generate requirement models and specifications, surpassing SOTA baselines using text parser and logical reasoning by 15.4\%.

\subsection{Code Generation}
Code generation has been a long-standing focus of SE research. A prominent multi-agent setup for coding typically relies on role specialization and iterative feedback loops to optimize collaboration among agents. Common roles include orchestrator, programmer, reviewer, tester, and information retriever.

\textbf{Code generation with planning}: 
Chain-of-Thought (CoT) is a popular strategy in code generation~\cite{zelikman2023parsel,le2023codechain,huang2023codecot}. For instance, CodeCoT~\cite{huang2023codecot} leverages CoT to break down the requirements into steps described in natural language and then convert them to code.
% planning
Some other works employ planning strategies~\cite{shinn2024reflexion, bairi2024codeplan, wang2024teaching, zhang2024pair}. 
As an example, PairCoder~\cite{zhang2024pair} incorporates two collaborative LLM agents. The navigator agent proposes solution plans, selects the optimal plan, and controls the next iteration round based on execution feedback. The driver agent follows its guidance and performs initial code generation, testing, and refinement.

\textbf{Code generation with iterative refinement}: 
Multi-agent systems can dynamically refine the generated code through iterative feedback from \textbf{models}~\cite{wu2024autogen,li2023camel,chen2023teaching,jiang2023selfevolve}, \textbf{tools}~\cite{zhang2024codeagent, yang2024intercode,zhang2023toolcoder,zhang2024codeagent,he2024cocost, wu2024autogen,shinn2024reflexion,wang2024intervenor,elhashemy2025bridging}, and \textbf{humans}~\cite{wang2023mint,mu2023clarifygpt}. For instance, the agent can receive peer-reflection from multiple models with specialized roles~\cite{wu2024autogen,li2023camel} or conduct self-refinement based on the previous output~\cite{chen2023teaching,jiang2023selfevolve}. The agent can also collect informative compilation errors from compiler and interpreter tools~\cite{zhang2024codeagent, yang2024intercode}, or even get access to external resources by applying retrieval or searching tools~\cite{zhang2023toolcoder,zhang2024codeagent,he2024cocost}. Furthermore, hybrid feedback from tool and model is possible -- receiving error messages from tools and then using LLMs' contextual
understanding to provide corresponding explanations, suggestions, and instructions~\cite{wu2024autogen,shinn2024reflexion,wang2024intervenor}. Considering the critical role of human in clarifying ambiguous requirements, several works incorporate human feedback into the code generation process~\cite{wang2023mint,mu2023clarifygpt}.

\subsection{Static Code Checking}
Static code checking refers to examining the quality of code
without executing it, which is an essential step to identify diverse code quality issues, such as bugs, vulnerabilities, or code smells before executing the tests extensively.

\textbf{Static bug detection}: 
Preliminary studies show that fine-tuning LLMs on code base or simply prompting them has demonstrated superior effectiveness in identifying bugs in code snippets~\cite{feng2024prompting, yuan2023evaluating}. From the perspective of multi-agent collaboration, an approach for vulnerability detection~\cite{mao2024multi} simulates the real-world code review process with multiple test and developer agents. GPTLENS~\cite{hu2023large} creates a synergistic framework, where several auditor agents randomly generate potential
vulnerabilities and their corresponding reasoning thoroughly, while the critic agents carefully review and score them based on specific criteria. LLIFT~\cite{li2024enhancing} combines LLM-based agentic systems with traditional static analysis tools. Based on the undecided bugs reported by the tool UBITect, LLIFT further leverages LLMs' capability in code comprehension and summarization to identify the UBI bugs in Linux kernels.

\textbf{Code review}: 
In code review, LLM-based agents mimic the real-world peer review procedure by including multiple agents as different code reviewers. CodeAgent~\cite{Tang2024CodeAgentCA} simulates a waterfall-like pipeline with four stages and sets up a code review team with six agents of different characters, including user, CEO, CPO, CTO, coder, and reviewer. ICAA~\cite{fan2023static} designs a multi-agent system to identify code-intention inconsistencies, where the context incubation agent gathers essential information from code repositories, the consistency checking agent analyzes this information to identify any discrepancies, and the report agent compiles the final report.

\subsection{Testing and Quality Assurance}
Creating tests in practice can be challenging, as test cases directly generated by LLMs can encounter compilation or execution errors. The tests must be both syntactically and semantically accurate, while also being sufficiently comprehensive to cover as many different states of the software under test as possible.

\textbf{Unit test}: 
LLM-based agents can eliminate unit test errors by iteratively collecting error messages and fixing the buggy test code~\cite{yuan2023no,schafer2023empirical}. Similarly, CoverUp~\cite{pizzorno2024coverup} refines its prompts iteratively to focus on the code snippets that lack coverage, thus achieving higher test coverage rates.

\textbf{System test}: 
WhiteFox~\cite{yang2024whitefox} encompasses two agents working together. An analysis agent examines the low-level source code and produces requirements on the high-level test programs that can trigger the optimization, while a generation agent crafts test programs based on summarized requirements. In terms of mobile applications, GPTDroid~\cite{liu2024make} automates the entire process by perceiving GUI page information, generating test scripts in the form of Q\&A, executing these scripts using tools, and receiving feedback from the application. Regarding web applications, given an API name, RESTSpecIT~\cite{decrop2024you} generates and mutates HTTP requests through a reflection loop. By sending these requests to the API endpoint, it analyzes the HTTP responses for inference and testing. The agent then uses valid requests as feedback to refine the mutations in each iteration.

\subsection{Debugging}
%Software debugging typically includes two phases: fault localization and program repair.
 
\textbf{Fault localization}: 
AgentFL~\cite{qin2024agentfl} scales up LLM-based fault localization to project-level code context through the synergy of multiple agents. AUTOFL~\cite{kang2023preliminary}, a single-agent system, enhances standalone LLMs with tool invocation, utilizing four specialized function calls to explore the repository more effectively.

\textbf{Program repair}: 
Feedback-driven approaches dominate recent program repair research. ChatRepair\cite{xia2024automated} refines patches based on iterative environmental feedback, while CigaR\cite{hidvegi2024cigar} applies similar principles to function-level repair. AutoSD~\cite{kang2025explainable} adopts a scientific debugging methodology, integrating hypothesis generation, execution-based validation, conclusion making, and automated fixing into a unified repair pipeline.

%ChatRepair~\cite{xia2024automated} iteratively refines patches based on environmental feedback. CigaR~\cite{hidvegi2024cigar} works similarly for function-level program repair. AutoSD~\cite{kang2025explainable} fixes the buggy program iteratively by simulating scientific debugging and including four components: hypothesis generator, execution-based validator, conclusion maker, and fixer.

\section{Methodological Approaches and Research Agenda}

\subsection{Model Selection and Evaluation}
The effectiveness of multi-agent systems fundamentally depends on the capabilities of the underlying language models. However, deploying such systems in real-world software engineering scenarios requires careful tradeoffs between model performance, computational cost, and data privacy. 

\textbf{Proprietary frontier models} represent performance upper bounds. Notable examples include GPT-5 \cite{gpt5} from OpenAI, Gemini-3-pro \cite{gemini-3-pro} from Google, and Claude-v4.5 \cite{claude} developed by Anthropic, which demonstrate forefront performance in language understanding and generation. However, these models require API-based access, which raises significant privacy and security concerns when processing proprietary codebases.

\textbf{Open-source alternatives} address privacy and cost constraints while enabling local deployment and customization. For instance, LLaMA-4 \cite{llama-4} and DeepSeek-V3 \cite{liu2024deepseek} can achieve competitive performance to proprietary models. Such open-source LLMs enable organizations to maintain full control over sensitive codebases, eliminate recurring API costs, and customize model behavior for domain-specific SE tasks, though requiring higher infrastructure investment.

\textbf{Reasoning-optimized architectures} represent a recent paradigm shift in model design. 
Compared with contemporaneous models, DeepSeek-R1 %~\cite{deepseekai2025deepseekr1incentivizingreasoningcapability} 
incorporates specialized reasoning capabilities, which yields advantages for complex software engineering tasks like debugging and architectural decision-making.
As new competent models emerge, evaluating them systematically across various SE tasks allows us to discover alternatives that balance cost-efficiency, privacy protection, and strong performance in addressing complex software engineering problems.
%As new competent models emerge continuously, systematic evaluation of diverse models across various SE tasks enables us to identify cost-effective and privacy-preserving alternatives that maintain competitive performance for complex SE challenges.

\subsection{Comparative Analysis and Benchmarks} 
The software engineering domain offers a rich ecosystem of open-source datasets and benchmarks spanning diverse tasks. For instance in requirements engineering, PURE~\cite{ferrari2017pure} and PROMISE~\cite{li2014non} datasets provide standardized evaluation data for software requirement classification. For code generation, humanEval~\cite{chen2021evaluating} emphasizes the functional correctness on programming problems, while GSM8K~\cite{cobbe2021gsm8k} assesses mathematical reasoning in code. With respect to quality assurance, BugBench~\cite{lu2005bugbench} provides a benchmark for bug detection. Tests4Py~\cite{smytzek2024tests4py} focuses on system testing scenarios. By evaluating LLM-based solutions across these established benchmarks, we provide empirical evidence for the effectiveness of our approaches while enabling direct comparison with existing methods.

\subsection{Agentic Frameworks and Coordination}
To facilitate the development and deployment of agentic solutions, the community has proposed a wide range of \textbf{frameworks for agents}, including but not limited to \textbf{CrewAI, LangGraph, Google ADK, AutoGen, and n8n}. The rapid growth of these frameworks highlights the need to evaluate their architectural principles, communication mechanisms, memory management strategies, and safety guardrails \cite{derouiche2025agenticaiframeworksarchitectures}. 

In parallel with these frameworks, \textbf{agent communication protocols} have been introduced to standardize inter-agent interactions and promote collaborative problem-solving \cite{yang2025survey}. Widely-used examples include Contract Net Protocol (CNP), Agent-to-Agent (A2A), and  Agent Network Protocol (ANP). Meanwhile, \textbf{Model Context Protocol (MCP)} has emerged as an open standard that unifies the integration of external tools and data sources.

Despite the abundance and diversity of these frameworks and protocols, deploying them in real-world software projects raises significant concerns on reliability and privacy. As development teams transition from proof-of-concept demonstrations to production environments, they might encounter critical challenges, including dynamic agent discovery, seamless integration of human feedback, and effective management of computational costs and data constraints. These practical deployment gaps motivate the investigation of key challenges and research opportunities in Section \ref{sec: challenges}. 

\section{Future Challenges and Opportunities} 
\label{sec: challenges}
Despite the promising future of LLM-based multi-agent systems throughout the Software Development Life Cycle (SDLC), numerous challenges remain:
\begin{itemize}
    \item \textbf{Enhancing individual agents' capabilities}: %We can strengthen individual agents from two perspectives -- refining role-playing and prompting strategies. 
    Current LLM-based agentic systems can only simulate generic roles like software developers and product managers. They still lack the nuanced expertise to mimic more SE-specific roles, such as for vulnerability detection or security auditing. %For example, multiple studies have identified ChatGPT's deficiency in detecting and repairing vulnerabilities. 
    This shortcoming emphasizes the need to integrate domain-specific knowledge into LLMs to support specialized SE roles. Meanwhile, integrating effective prompting frameworks can significantly facilitate agentic systems. Investigating more efficient Retrieval Augmented Generation (RAG) \cite{gao2023retrieval} techniques and agent-based workflows will remain a prominent area of research in the future.
    
    \item \textbf{Optimizing agent synergy and human-agent coordination}:
    Given the diverse strength of individual agents, the next phase of research focuses on improving the internal dynamics among agents and the external human interaction with multi-agent systems. While some studies have incorporated human-in-the-loop designs, e.g., MARE system \cite{jin2024mare} for requirement engineering, many research questions remain unanswered. These include optimizing human roles, enhancing feedback mechanisms, and identifying appropriate intervention points in human-agent collaboration.
    
    \item \textbf{Collecting more data throughout the SDLC}: Most agents rely on general-purpose LLMs such as ChatGPT or code-specific models like Deepseek-Coder. However, complex SE tasks require more specific data beyond code. Collecting and utilizing valuable data from the entire SDLC, such as design, architecture, developer discussions, historical code changes, and dynamic runtime information, can enhance LLMs' capabilities significantly.

    \item \textbf{Reducing computational efforts and costs}: Reducing the costs associated with deploying and maintaining LLM-based multi-agent systems is crucial for ensuring scalability, sustainability, and cost-effectiveness in real-world projects. Distilling large-scale models or fine-tuning light-weight models for a specific SE agent could be a solution; incorporating robust feedback mechanisms and minimizing iterations between agents can also lower the computational costs for LLM inference.

    \item \textbf{Evaluating multi-agent systems for SE}: Existing benchmarks have focused on individual SE tasks and exhibit limitations as SE projects grow in complexity. However, software engineering is inherently collaborative, including activities like joint requirements gathering, code integration, and peer reviews. Current benchmarks often overlook these aspects, highlighting an increasing need to assess LLMs' cooperative capabilities within multi-agent settings.

 \end{itemize}

\section{Conclusion}
%In this work, we review the most recent advancements in leveraging multi-agent systems for software engineering, propose potential research directions, and discuss the future challenges in this field. To move forward, we appeal to more attention and an open discussion in the research community, proactively applying LLM-based agentic systems to more practical applications throughout the entire software development process. 

% In conclusion, this work has reviewed the latest advancements in leveraging LLM-based multi-agent systems for software engineering, highlighting their potential impact across various stages in SDLC. While these systems demonstrate significant promise, we emphasize that achieving fully automated software development demands more than success in isolated tasks. 

In this work, we reviewed the most recent advancements in LLM-based multi-agent systems for SE, proposed potential research directions, and discussed future challenges for practical deployment. Our analysis reveals their promise across various SDLC stages, but also emphasizes that fully automated software development requires progress beyond individual stage. Software engineering is inherently collaborative, involving collective decision-making, iterative peer refinement, and continuous coordination among diverse roles.

As for future work, We plan to compare multi-agent design patterns, such as planner–executor pairs, self-refinement loops, and human-in-the-loop, and perform empirical evaluations on more open-source and proprietary LLMs. To move forward and exploit the full potential of agentic systems for SE, we encourage the entire community to work jointly and overcome current limitations together.

\section*{Acknowledgments} % Acknowledgment section
This work was carried out within the ITEA 4 project GENIUS, as part of the ITEA programme, the Eureka Cluster on software innovation. This work was funded by the German Federal Ministry of Research, Technology and Space (BMFTR).

\printbibliography

@misc{zhao2024surveylargelanguagemodels,
      title={A Survey of Large Language Models}, 
      author={Wayne Xin Zhao and Kun Zhou and Junyi Li and Tianyi Tang and Xiaolei Wang and Yupeng Hou and Yingqian Min and Beichen Zhang and Junjie Zhang and Zican Dong and Yifan Du and Chen Yang and Yushuo Chen and Zhipeng Chen and Jinhao Jiang and Ruiyang Ren and Yifan Li and Xinyu Tang and Zikang Liu and Peiyu Liu and Jian-Yun Nie and Ji-Rong Wen},
      year={2024},
      eprint={2303.18223},
      archivePrefix={arXiv},
      primaryClass={cs.CL},
      url={https://arxiv.org/abs/2303.18223}, 
}

@article{vaswani2017attention,
  title={Attention is all you need},
  author={Vaswani, Ashish and Shazeer, Noam and Parmar, Niki and Uszkoreit, Jakob and Jones, Llion and Gomez, Aidan N and Kaiser, {\L}ukasz and Polosukhin, Illia},
  journal={Advances in Neural Information Processing Systems},
  volume={30},
  year={2017}
}

@inproceedings{Devlin2019BERTPO,
  title={BERT: Pre-training of Deep Bidirectional Transformers for Language Understanding},
  author={Jacob Devlin and Ming-Wei Chang and Kenton Lee and Kristina Toutanova},
  booktitle={North American Chapter of the Association for Computational Linguistics},
  year={2019},
  url={https://api.semanticscholar.org/CorpusID:52967399}
}

@misc{wei2022emergent,
      title={Emergent Abilities of Large Language Models}, 
      author={Jason Wei and Yi Tay and Rishi Bommasani and Colin Raffel and Barret Zoph and Sebastian Borgeaud and Dani Yogatama and Maarten Bosma and Denny Zhou and Donald Metzler and Ed H. Chi and Tatsunori Hashimoto and Oriol Vinyals and Percy Liang and Jeff Dean and William Fedus},
      year={2022},
      eprint={2206.07682},
      archivePrefix={arXiv},
      primaryClass={cs.CL}
}

@inproceedings{sami2024early,
  title={Early Results of an AI Multiagent System for Requirements Elicitation and Analysis},
  author={Sami, Malik Abdul and Waseem, Muhammad and Zhang, Zheying and Rasheed, Zeeshan and Syst{\"a}, Kari and Abrahamsson, Pekka},
  booktitle={International Conference on Product-Focused Software Process Improvement},
  pages={307--316},
  year={2024},
  organization={Springer}
}

@article{ma2024specgen,
  title={SpecGen: Automated Generation of Formal Program Specifications via Large Language Models},
  author={Ma, Lezhi and Liu, Shangqing and Li, Yi and Xie, Xiaofei and Bu, Lei},
  journal={arXiv preprint arXiv:2401.08807},
  year={2024}
}

@article{jin2024mare,
  title={MARE: Multi-Agents Collaboration Framework for Requirements Engineering},
  author={Jin, Dongming and Jin, Zhi and Chen, Xiaohong and Wang, Chunhui},
  journal={arXiv preprint arXiv:2405.03256},
  year={2024}
}

@article{zelikman2023parsel,
  title={Parsel: Algorithmic Reasoning with Language Models by Composing Decompositions},
  author={Zelikman, Eric and Huang, Qian and Poesia, Gabriel and Goodman, Noah and Haber, Nick},
  journal={Advances in Neural Information Processing Systems},
  volume={36},
  pages={31466--31523},
  year={2023}
}

@article{le2023codechain,
  title={Codechain: Towards modular code generation through chain of self-revisions with representative sub-modules},
  author={Le, Hung and Chen, Hailin and Saha, Amrita and Gokul, Akash and Sahoo, Doyen and Joty, Shafiq},
  journal={arXiv preprint arXiv:2310.08992},
  year={2023}
}

@article{huang2023codecot,
  title={Codecot: Tackling code syntax errors in cot reasoning for code generation},
  author={Huang, Dong and Bu, Qingwen and Qing, Yuhao and Cui, Heming},
  journal={CoRR},
  volume={2308},
  pages={1--20},
  year={2023}
}

@article{shinn2024reflexion,
  title={Reflexion: Language agents with verbal reinforcement learning},
  author={Shinn, Noah and Cassano, Federico and Gopinath, Ashwin and Narasimhan, Karthik and Yao, Shunyu},
  journal={Advances in Neural Information Processing Systems},
  volume={36},
  year={2024}
}

@article{bairi2024codeplan,
  title={Codeplan: Repository-level coding using llms and planning},
  author={Bairi, Ramakrishna and Sonwane, Atharv and Kanade, Aditya and Iyer, Arun and Parthasarathy, Suresh and Rajamani, Sriram and Ashok, B and Shet, Shashank},
  journal={Proceedings of the ACM on Software Engineering},
  volume={1},
  number={FSE},
  pages={675--698},
  year={2024},
  publisher={ACM New York, NY, USA}
}

@article{wang2024teaching,
  title={Teaching Code LLMs to Use Autocompletion Tools in Repository-Level Code Generation},
  author={Wang, Chong and Zhang, Jian and Feng, Yebo and Li, Tianlin and Sun, Weisong and Liu, Yang and Peng, Xin},
  journal={arXiv preprint arXiv:2401.06391},
  year={2024}
}

@inproceedings{zhang2024pair,
  title={A Pair Programming Framework for Code Generation via Multi-Plan Exploration and Feedback-Driven Refinement},
  author={Zhang, Huan and Cheng, Wei and Wu, Yuhan and Hu, Wei},
  booktitle={Proceedings of the 39th IEEE/ACM International Conference on Automated Software Engineering},
  pages={1319--1331},
  year={2024}
}

@article{li2023camel,
  title={Camel: Communicative agents for" mind" exploration of large language model society},
  author={Li, Guohao and Hammoud, Hasan and Itani, Hani and Khizbullin, Dmitrii and Ghanem, Bernard},
  journal={Advances in Neural Information Processing Systems},
  volume={36},
  pages={51991--52008},
  year={2023}
}

@inproceedings{wu2024autogen,
  title={Autogen: Enabling next-gen LLM applications via multi-agent conversations},
  author={Wu, Qingyun and Bansal, Gagan and Zhang, Jieyu and Wu, Yiran and Li, Beibin and Zhu, Erkang and Jiang, Li and Zhang, Xiaoyun and Zhang, Shaokun and Liu, Jiale and others},
  booktitle={First Conference on Language Modeling},
  year={2024}
}

@article{chen2023teaching,
  title={Teaching large language models to self-debug},
  author={Chen, Xinyun and Lin, Maxwell and Sch{\"a}rli, Nathanael and Zhou, Denny},
  journal={arXiv preprint arXiv:2304.05128},
  year={2023}
}

@article{jiang2023selfevolve,
  title={Selfevolve: A code evolution framework via large language models},
  author={Jiang, Shuyang and Wang, Yuhao and Wang, Yu},
  journal={arXiv preprint arXiv:2306.02907},
  year={2023}
}

@article{zhang2024codeagent,
  title={Codeagent: Enhancing code generation with tool-integrated agent systems for real-world repo-level coding challenges},
  author={Zhang, Kechi and Li, Jia and Li, Ge and Shi, Xianjie and Jin, Zhi},
  journal={arXiv preprint arXiv:2401.07339},
  year={2024}
}

@article{yang2024intercode,
  title={Intercode: Standardizing and benchmarking interactive coding with execution feedback},
  author={Yang, John and Prabhakar, Akshara and Narasimhan, Karthik and Yao, Shunyu},
  journal={Advances in Neural Information Processing Systems},
  volume={36},
  year={2024}
}

@article{zhang2023toolcoder,
  title={Toolcoder: Teach code generation models to use api search tools},
  author={Zhang, Kechi and Zhang, Huangzhao and Li, Ge and Li, Jia and Li, Zhuo and Jin, Zhi},
  journal={arXiv preprint arXiv:2305.04032},
  year={2023}
}

@article{he2024cocost,
  title={CoCoST: Automatic Complex Code Generation with Online Searching and Correctness Testing},
  author={He, Xinyi and Zou, Jiaru and Lin, Yun and Zhou, Mengyu and Han, Shi and Yuan, Zejian and Zhang, Dongmei},
  journal={arXiv preprint arXiv:2403.13583},
  year={2024}
}

@inproceedings{wang2024intervenor,
  title={Intervenor: Prompting the coding ability of large language models with the interactive chain of repair},
  author={Wang, Hanbin and Liu, Zhenghao and Wang, Shuo and Cui, Ganqu and Ding, Ning and Liu, Zhiyuan and Yu, Ge},
  booktitle={Findings of the Association for Computational Linguistics ACL 2024},
  pages={2081--2107},
  year={2024}
}

@article{wang2023mint,
  title={Mint: Evaluating llms in multi-turn interaction with tools and language feedback},
  author={Wang, Xingyao and Wang, Zihan and Liu, Jiateng and Chen, Yangyi and Yuan, Lifan and Peng, Hao and Ji, Heng},
  journal={arXiv preprint arXiv:2309.10691},
  year={2023}
}

@article{mu2023clarifygpt,
  title={ClarifyGPT: Empowering LLM-based Code Generation with Intention Clarification},
  author={Mu, Fangwen and Shi, Lin and Wang, Song and Yu, Zhuohao and Zhang, Binquan and Wang, Chenxue and Liu, Shichao and Wang, Qing},
  journal={arXiv preprint arXiv:2310.10996},
  year={2023}
}

@inproceedings{feng2024prompting,
  title={Prompting is all you need: Automated android bug replay with large language models},
  author={Feng, Sidong and Chen, Chunyang},
  booktitle={Proceedings of the 46th IEEE/ACM International Conference on Software Engineering},
  pages={1--13},
  year={2024}
}

@article{yuan2023evaluating,
  title={Evaluating instruction-tuned large language models on code comprehension and generation},
  author={Yuan, Zhiqiang and Liu, Junwei and Zi, Qiancheng and Liu, Mingwei and Peng, Xin and Lou, Yiling},
  journal={arXiv preprint arXiv:2308.01240},
  year={2023}
}

@inproceedings{mao2024multi,
  title={Multi-role consensus through llms discussions for vulnerability detection},
  author={Mao, Zhenyy and Li, Jialong and Jin, Dongming and Li, Munan and Tei, Kenji},
  booktitle={2024 IEEE 24th International Conference on Software Quality, Reliability, and Security Companion (QRS-C)},
  pages={1318--1319},
  year={2024},
  organization={IEEE}
}

@inproceedings{hu2023large,
  title={Large language model-powered smart contract vulnerability detection: New perspectives},
  author={Hu, Sihao and Huang, Tiansheng and {\.I}lhan, Fatih and Tekin, Selim Furkan and Liu, Ling},
  booktitle={2023 5th IEEE International Conference on Trust, Privacy and Security in Intelligent Systems and Applications (TPS-ISA)},
  pages={297--306},
  year={2023},
  organization={IEEE}
}

@article{li2024enhancing,
  title={Enhancing Static Analysis for Practical Bug Detection: An LLM-Integrated Approach},
  author={Li, Haonan and Hao, Yu and Zhai, Yizhuo and Qian, Zhiyun},
  journal={Proceedings of the ACM on Programming Languages},
  volume={8},
  number={OOPSLA1},
  pages={474--499},
  year={2024},
  publisher={ACM New York, NY, USA}
}

@article{Tang2024CodeAgentCA,
  title={CodeAgent: Collaborative Agents for Software Engineering},
  author={Daniel Tang and Zhenghan Chen and Kisub Kim and Yewei Song and Haoye Tian and Saad Ezzini and Yongfeng Huang and Jacques Klein and T{\'e}gawend{\'e} F. Bissyand{\'e}},
  journal={ArXiv},
  year={2024},
  volume={abs/2402.02172},
  url={https://api.semanticscholar.org/CorpusID:270865202}
}

@article{fan2023static,
  title={Static Code Analysis in the AI Era: An In-depth Exploration of the Concept, Function, and Potential of Intelligent Code Analysis Agents},
  author={Fan, Gang and Xie, Xiaoheng and Zheng, Xunjin and Liang, Yinan and Di, Peng},
  journal={arXiv preprint arXiv:2310.08837},
  year={2023}
}

@article{yuan2023no,
  title={No more manual tests? evaluating and improving chatgpt for unit test generation},
  author={Yuan, Zhiqiang and Lou, Yiling and Liu, Mingwei and Ding, Shiji and Wang, Kaixin and Chen, Yixuan and Peng, Xin},
  journal={arXiv preprint arXiv:2305.04207},
  year={2023}
}

@article{schafer2023empirical,
  title={An empirical evaluation of using large language models for automated unit test generation},
  author={Sch{\"a}fer, Max and Nadi, Sarah and Eghbali, Aryaz and Tip, Frank},
  journal={IEEE Transactions on Software Engineering},
  year={2023},
  publisher={IEEE}
}

@article{pizzorno2024coverup,
  title={CoverUp: Coverage-Guided LLM-Based Test Generation},
  author={Pizzorno, Juan Altmayer and Berger, Emery D},
  journal={arXiv preprint arXiv:2403.16218},
  year={2024}
}

@article{yang2024whitefox,
  title={Whitefox: White-box compiler fuzzing empowered by large language models},
  author={Yang, Chenyuan and Deng, Yinlin and Lu, Runyu and Yao, Jiayi and Liu, Jiawei and Jabbarvand, Reyhaneh and Zhang, Lingming},
  journal={Proceedings of the ACM on Programming Languages},
  volume={8},
  number={OOPSLA2},
  pages={709--735},
  year={2024},
  publisher={ACM New York, NY, USA}
}

@inproceedings{liu2024make,
  title={Make llm a testing expert: Bringing human-like interaction to mobile gui testing via functionality-aware decisions},
  author={Liu, Zhe and Chen, Chunyang and Wang, Junjie and Chen, Mengzhuo and Wu, Boyu and Che, Xing and Wang, Dandan and Wang, Qing},
  booktitle={Proceedings of the IEEE/ACM 46th International Conference on Software Engineering},
  pages={1--13},
  year={2024}
}

@article{decrop2024you,
  title={You Can REST Now: Automated Specification Inference and Black-Box Testing of RESTful APIs with Large Language Models},
  author={Decrop, Alix and Perrouin, Gilles and Papadakis, Mike and Devroey, Xavier and Schobbens, Pierre-Yves},
  journal={arXiv preprint arXiv:2402.05102},
  year={2024}
}

@article{qin2024agentfl,
  title={AgentFL: Scaling LLM-based Fault Localization to Project-Level Context},
  author={Qin, Yihao and Wang, Shangwen and Lou, Yiling and Dong, Jinhao and Wang, Kaixin and Li, Xiaoling and Mao, Xiaoguang},
  journal={arXiv preprint arXiv:2403.16362},
  year={2024}
}

@article{kang2023preliminary,
  title={A preliminary evaluation of llm-based fault localization},
  author={Kang, Sungmin and An, Gabin and Yoo, Shin},
  journal={arXiv preprint arXiv:2308.05487},
  year={2023}
}

@inproceedings{xia2024automated,
  title={Automated program repair via conversation: Fixing 162 out of 337 bugs for {\$}0.42 each using chatgpt},
  author={Xia, Chunqiu Steven and Zhang, Lingming},
  booktitle={Proceedings of the 33rd ACM SIGSOFT International Symposium on Software Testing and Analysis},
  pages={819--831},
  year={2024}
}

@article{hidvegi2024cigar,
  title={Cigar: Cost-efficient program repair with llms},
  author={Hidv{\'e}gi, D{\'a}vid and Etemadi, Khashayar and Bobadilla, Sofia and Monperrus, Martin},
  journal={arXiv preprint arXiv:2402.06598},
  year={2024}
}

@article{kang2025explainable,
  title={Explainable automated debugging via large language model-driven scientific debugging},
  author={Kang, Sungmin and Chen, Bei and Yoo, Shin and Lou, Jian-Guang},
  journal={Empirical Software Engineering},
  volume={30},
  number={2},
  pages={1--28},
  year={2025},
  publisher={Springer}
}

@article{liu2024deepseek,
  title={Deepseek-v3 technical report},
  author={Liu, Aixin and Feng, Bei and Xue, Bing and Wang, Bingxuan and Wu, Bochao and Lu, Chengda and Zhao, Chenggang and Deng, Chengqi and Zhang, Chenyu and Ruan, Chong and others},
  journal={arXiv preprint arXiv:2412.19437},
  year={2024}
}

@inproceedings{li2014non,
  title={Non-functional requirements as qualities, with a spice of ontology},
  author={Li, Feng-Lin and Horkoff, Jennifer and Mylopoulos, John and Guizzardi, Renata SS and Guizzardi, Giancarlo and Borgida, Alexander and Liu, Lin},
  booktitle={2014 IEEE 22nd International Requirements Engineering Conference (RE)},
  pages={293--302},
  year={2014},
  organization={IEEE}
}

@inproceedings{ferrari2017pure,
  title={Pure: A dataset of public requirements documents},
  author={Ferrari, Alessio and Spagnolo, Giorgio Oronzo and Gnesi, Stefania},
  booktitle={2017 IEEE 25th international requirements engineering conference (RE)},
  pages={502--505},
  year={2017},
  organization={IEEE}
}

@article{cobbe2021gsm8k,
  title={Training Verifiers to Solve Math Word Problems},
  author={Cobbe, Karl and Kosaraju, Vineet and Bavarian, Mohammad and Chen, Mark and Jun, Heewoo and Kaiser, Lukasz and Plappert, Matthias and Tworek, Jerry and Hilton, Jacob and Nakano, Reiichiro and Hesse, Christopher and Schulman, John},
  journal={arXiv preprint arXiv:2110.14168},
  year={2021}
}

@misc{chen2021evaluating,
      title={Evaluating Large Language Models Trained on Code},
      author={Mark Chen and Jerry Tworek and Heewoo Jun and Qiming Yuan and Henrique Ponde de Oliveira Pinto and Jared Kaplan and Harri Edwards and Yuri Burda and Nicholas Joseph and others},
      year={2021},
      eprint={2107.03374},
      archivePrefix={arXiv},
      primaryClass={cs.LG}
}

@inproceedings{lu2005bugbench,
  title={Bugbench: Benchmarks for evaluating bug detection tools},
  author={Lu, Shan and Li, Zhenmin and Qin, Feng and Tan, Lin and Zhou, Pin and Zhou, Yuanyuan},
  booktitle={Workshop on the evaluation of software defect detection tools},
  volume={5},
  year={2005},
  organization={Chicago, Illinois}
}

@inproceedings{smytzek2024tests4py,
  title={Tests4Py: A Benchmark for System Testing},
  author={Smytzek, Marius and Eberlein, Martin and Serce, Batuhan and Grunske, Lars and Zeller, Andreas},
  booktitle={Companion Proceedings of the 32nd ACM International Conference on the Foundations of Software Engineering},
  pages={557--561},
  year={2024}
}

@incollection{tang2024fsponer,
  title={FsPONER: Few-Shot Prompt Optimization for Named Entity Recognition in Domain-Specific Scenarios},
  author={Tang, Yongjian and Hasan, Rakebul and Runkler, Thomas},
  booktitle={ECAI 2024},
  pages={3757--3764},
  year={2024},
  publisher={IOS Press}
}

@article{sahoo2024systematic,
  title={A systematic survey of prompt engineering in large language models: Techniques and applications},
  author={Sahoo, Pranab and Singh, Ayush Kumar and Saha, Sriparna and Jain, Vinija and Mondal, Samrat and Chadha, Aman},
  journal={arXiv preprint arXiv:2402.07927},
  year={2024}
}

@inproceedings{dong2024survey,
  title={A survey on in-context learning},
  author={Dong, Qingxiu and Li, Lei and Dai, Damai and Zheng, Ce and Ma, Jingyuan and Li, Rui and Xia, Heming and Xu, Jingjing and Wu, Zhiyong and Chang, Baobao and others},
  booktitle={Proceedings of the 2024 conference on empirical methods in natural language processing},
  pages={1107--1128},
  year={2024}
}

@article{hanparameter,
  title={Parameter-Efficient Fine-Tuning for Large Models: A Comprehensive Survey},
  author={Han, Zeyu and Gao, Chao and Liu, Jinyang and Zhang, Jeff and Zhang, Sai Qian},
  journal={Transactions on Machine Learning Research},
year = {2024}
}

@article{gao2023retrieval,
  title={Retrieval-augmented generation for large language models: A survey},
  author={Gao, Yunfan and Xiong, Yun and Gao, Xinyu and Jia, Kangxiang and Pan, Jinliu and Bi, Yuxi and Sun, Jiawei and Wang, Haofen},
year = {2023}

}

@misc{gpt5,
  author = {OpenAI},
  title = {{GPT-5} System Card},
  year = {2025},
  note = {Accessed August 7},
  url = {https://openai.com/index/gpt-5-system-card/}
}

@misc{claude,
  author = {Anthropic},
  title = {Introducing {Claude Sonnet 4.5d}},
  year = {2025},
  note = {Accessed September 29},
  url = {https://www.anthropic.com/news/claude-sonnet-4-5}
}

@misc{gemini-3-pro,
      author = {Google},
  title = {Gemini 3 Pro - Best for complex tasks and bringing creative concepts to life},
  year = {2025},
  url = {https://deepmind.google/models/gemini/pro/}
}

@misc{llama-4,
    author = {Meta},
  title = {{LLaMA 4}: Leading multimodal intelligence},
  year = {2025},
  url = {https://ai.meta.com/blog/llama-4-multimodal-intelligence/}
}

@misc{derouiche2025agenticaiframeworksarchitectures,
      title={Agentic AI Frameworks: Architectures, Protocols, and Design Challenges}, 
      author={Hana Derouiche and Zaki Brahmi and Haithem Mazeni},
      year={2025},
      eprint={2508.10146},
      archivePrefix={arXiv},
      primaryClass={cs.AI},
      url={https://arxiv.org/abs/2508.10146}, 
}

@article{yang2025survey,
  title={A survey of ai agent protocols},
  author={Yang, Yingxuan and Chai, Huacan and Song, Yuanyi and Qi, Siyuan and Wen, Muning and Li, Ning and Liao, Junwei and Hu, Haoyi and Lin, Jianghao and Chang, Gaowei and others},
  journal={arXiv preprint arXiv:2504.16736},
  year={2025}
}

@article{tang2025few,
  title={The few-shot dilemma: Over-prompting large language models},
  author={Tang, Yongjian and Tuncel, Doruk and Koerner, Christian and Runkler, Thomas},
  journal={arXiv preprint arXiv:2509.13196},
  year={2025}
}

@INPROCEEDINGS{nishtha,
  author={Vaidya, Nishtha N. and Runkler, Thomas A. and Hubauer, Thomas and Haderlein-Hoegberg, Veronika and Brandt, Maja Milicic},
  booktitle={2025 IEEE Symposium on Computational Intelligence in Natural Language Processing and Social Media (CI-NLPSoMe)}, 
  title={Conceptual In-Context Learning and Chain of Concepts: Solving Complex Conceptual Problems Using Large Language Models}, 
  year={2025},
  volume={},
  number={},
  pages={1-7},

  doi={10.1109/CI-NLPSoMe64976.2025.10970773}}

@article{elhashemy2025bridging,
  title={Bridging the Prototype-Production Gap: A Multi-Agent System for Notebooks Transformation},
  author={Elhashemy, Hanya and Lotfy, Youssef and Tang, Yongjian},
  journal={arXiv preprint arXiv:2511.07257},
  year={2025}
}
%\bibliography{ref}

\end{document}